# Explainable AI for Enhancing IDS Against Advanced Persistent Kill Chain

Bassam Noori Shaker[1,2*] , Bahaa Al-Musawi[3] , Mohammed Falih Hassan[3]

[1*]Faculty of Computer Science and Mathematics, University of Kufa, Al-Najaf, Iraq.
[2]Faculty of Computer Science and Information Technology, University of Al-Qadisiyah, Iraq
[3]Department of Electronic and Communications, Faculty of Engineering, University of Kufa, Al-Najaf, Iraq.

**Abstract.** Advanced Persistent Threats (APTs) represent a sophisticated and persistent cybersecurity challenge, characterized by stealthy, multi-phase, and targeted attacks aimed at compromising information systems over an extended period. Developing an effective Intrusion Detection System (IDS) capable of detecting APTs at different phases relies on selecting network traffic features. However, not all of these features are directly related to the phases of APTs. Some network traffic features may be unrelated or have limited relevance to identifying malicious activity. Therefore, it is important to carefully select and analyze the most relevant features to improve the IDS performance. This work proposes a feature selection and classification model that integrates two prominent machine learning algorithms: SHapley Additive exPlanations (SHAP) and Extreme Gradient Boosting (XGBoost). The aim is to develop lightweight IDS based on a selected minimum number of influential features for detecting APTs at various phases. The proposed method also specifies the relevant features for each phase of APTs independently. Extensive experimental results on the SCVIC-APT-2021 dataset indicated that our proposed approach has improved performance compared to other standard techniques. Specifically, both the macro-average F1-score and recall reached 94% and 93 %, respectively, while reducing the complexity of the detection model by selecting only 12 features out of 77.

## 1      Introduction

Advanced Persistent Threat (APT) is a multi-layered cyber-attack characterized by sophistication and a prolonged duration [1]. APTs often remain undetected for extended periods while exfiltrating sensitive data or causing severe disruptions. An APT follows a series of phases known as the kill chain, where an attacker must complete each phase to reach their objective. Each phase includes one or more classes of threats or attacks. The APT protection market is expected to grow to over 23 billion U.S. dollars by 2028, up from 10 billion U.S. dollars in 2024. This significant growth highlights the increasing need for robust cybersecurity tools for detecting APTs such as Intrusion Detection Systems (IDSs), Endpoint Protection Platforms (EPPs), and Threat Intelligence Platforms (TIPs) [2].

An IDS is a security tool designed to monitor network traffic to spot any malicious activities or breaches of security policies. By scrutinizing network and system data, an IDS looks for unusual patterns that may signal potential attacks or unauthorized access [3]. IDS models are developed using machine learning (ML) algorithms that enable the system to learn how to distinguish between normal and malicious network traffic. These algorithms are typically referred to as classification algorithms because they categorize incoming data into different classes, such as 'normal' or 'malicious.' By training on historical traffic data, the IDS can identify patterns and anomalies that indicate potential threats, making it an essential tool for maintaining network security [4] [5].

To produce a dependable IDS model capable of accurately detecting anomalies, the quality and composition of the dataset are crucial. A well-curated dataset ensures that the IDS can effectively identify deviations from normal behavior, leading to more reliable anomaly detection. The dataset must consist of a set of features, which are specific data attributes that serve as input to the classification algorithm. These features could include various characteristics of network traffic or system behavior, such as packet size, traffic volume, protocol types, connection duration, and user activity patterns. The classification algorithm uses these features to learn and identify patterns associated with both normal operations and security threats.

The input features for IDS models come from network traffic flow, often extracted using specialized tools like CICFlowMeter [6]. These tools convert raw network traffic stored in PCAP files (Packet Capture) into CSV files. For developing reliable IDS that can detect advanced threats like APT, it is preferable to train and evaluate IDS using a specialized APT dataset, with two important characteristics. Firstly, the datasets should include labelled samples representing various phases of an APT. This allows researchers to identify and distinguish between different phases of an APT, which is more effective than simply classifying traffic as "normal" or "malicious". For example, a dataset can contain different APT phases of traffic. By training a model to recognize these distinct phases, it becomes possible to detect not only the presence of an attack but also its progression, enabling more targeted and timely responses to each phase of the APT [7]. This, in turn, facilitates the development of more sophisticated and adaptive IDS solutions capable of effectively detecting and mitigating APTs. Secondly, the dataset should have fewer but more.

relevant features. While many features can provide insights into network activity, not all are equally useful for distinguishing between normal and malicious traffic. Therefore, strategically selecting the most pertinent features is crucial. This not only enhances IDS performance but also reduces training time, prevents overfitting, and strengthens the overall robustness of the IDS models [8]. Furthermore, reducing the number of features in an IDS can lead to a more lightweight and efficient system, especially in resource-constrained environments like IoT networks. Lightweight IDS op-erate effectively with minimal computational resources, such as memory and processing power. This is because they require less data to process, resulting in faster decision-making and lower latency in threat detection.

The feature selection method is mainly divided into two approaches: model-specific and model-agnostic. In the first approach, the method is closely linked to the internals of the model. The relevance or importance of a feature is determined by the model's structure and particular learning algorithm. Examples include coefficients in linear models (such as Lasso and Ridge) and feature importance scores, like those generated by Random Forest (RF) [9].

Model-agnostic approaches consider the ML model as a black box, meaning they can be applied to any kind of model. The advantage of these approaches is that they are faster to implement since they do not require interaction with the model to assess feature importance. However, analyzing features can help make the inner workings of the black box more transparent and understandable. Explainable Artificial Intelligence (XAI) methods, specifically SHapley Additive exPlanations (SHAP), can be used to explain the output of the ML model by analyzing the features. This method relies on SHAP values from game theory. In this framework, each feature is treated as a player, and the goal is to estimate the contribution of each feature toward the predicted outcomes. SHAP values compare the model's prediction with and without each player to determine their impact on the game [10]. SHAP provides two levels of explanation: local explanations, which determine what features matter for a particular individual prediction, and global explanations, which provide insights into how the model makes predictions and the relative importance of different features in the prediction process [11].

Recent IDS models are often developed based on advanced ML techniques to enhance their detection capabilities. One such technique is XGBoost, a highly effective gradient boosting algorithm that has demonstrated superior performance in intrusion detection tasks. XGBoost excels at handling large-scale datasets and capturing complex patterns within the data, making it particularly suitable for the dynamic and evolving nature of network security. As a powerful ensemble learning technique, XGBoost is well-suited for managing large, complex datasets, such as the vast amounts of network traffic data generated in modern computing environments. It builds a sequence of weak predictive models iteratively and combines them into a strong predictive model.

In this study, we employ SHAP, a robust and model-agnostic method for understanding feature importance, alongside the XGBoost algorithm to introduce a novel feature selection method. This method efficiently develops IDS models, preserving model accuracy while simplifying and reducing computational costs for data analysis.

Our contribution to this study involves:

- Proposing a hybrid feature selection system that combines the strength part of the SHAP analysis and XGBoost learning method.
- Determining the global features related to APT phases, as well as the local features relevant to each phase independently.
- Developing a lightweight IDS based on the minimum number of relevant features for detecting APTs at various phases.

The rest of this paper is organized as follows: In Section 2, we provide an overview of related works on detecting APT attacks. The main characteristics of APTs and the details of APT phases are introduced in Section 3. The methodology is outlined in Section 4. Section 5 presents the proposed feature selection method. Experimental setup and data preprocessing are declared in section 6. Results and discusses our study, focusing on the main findings present in Section 7, we summarize the main conclusions and contributions of our research in Section 8.

## 2  Related Works

In this section, we provide an overview of current studies related to APT detection, focusing on the methodologies and ML models.

Joloudari et al. in [12] proposed a DL model with six layers to extract and select APT features from the hidden layers of the neural network. This model was compared with the C5.0 decision tree and Bayesian classification models using the NSL-KDD dataset —an enhanced version of KDD-CUP, where duplicated records were removed and minority samples were increased—for evaluation [13]. Although the proposed model demonstrated the highest accuracy at 98.85% and the lowest false positive rate at 1.13%, established it as the preferred choice, it relies on the NSL-KDD dataset, which may not fully capture the complexity of real-world APT attacks that often involve more sophisticated, multi-stage techniques.

In [14], Shang et al. proposed a model to detect unknown APTs by analyzing network flow features in the communication channel between APT attacks and the C&C server, which attackers use to control compromised systems in a targeted network. This study employed two deep neural networks: a Long Short-Term Memory (LSTM) network to extract time series features at the packet level and a Convolutional Neural Network (CNN) to extract flow-based features. These features were then combined and reduced using the Principle Component Analysis (PCA) dimensionality reduction method. In the final step, different classifiers were used to detect the C&C channel of unknown APT attacks. While the study achieved an F1-score of 96%, it did not specifically detect individual phases of the APT lifecycle. Additionally, the model's complexity increased due to the incorporation of three different models for feature selection.

Martín Liras et al. in [15] proposed an approach for identifying the most discriminatory features that can distinguish APT-related malware from non-APT malware executables. They recommend using features from static, dynamic, and network-related analyses to achieve this goal. The study successfully identified 238 out of 1941 features related to APT attacks using three different feature selection methods: variance, the $\chi^2$ statistical test, and a tree-based estimator. The dataset with the selected features was evaluated for its accuracy in classifying APT-related malware using different ML algorithms, including logistic regression, Support Vector Machines (SVM), K-nearest neighbor (KNN), and RF. Although the experiment with the RF classifier achieved impressive results (an F1-score of 89% and an accuracy of 98%), the study did not replicate real-world scenarios by distinguishing between APT and normal traffic.

Due to the extended duration required to execute an APT attack, Yu, Keping et al. [16] leveraged this characteristic for the detection process in an Industrial Internet of Things (IIoT) environment by analyzing the time series of each phase. Each attack phase requires different time series data. They used a Bidirectional Encoder Representations from Transformers (BERT) model, which is a DL model specifically designed to capture long attack sequences and extended attack durations characteristic of APT attacks. The BERT model was compared with three other models: Perceptron, LSTM, and CNN. The results demonstrated that the BERT model outperformed the other models in terms of accuracy, achieving a rate as high as 99%. Although the study categorized the attacks into four levels based on sequence length, this approach does not accurately represent the actual phases of an APT.

Do Xuan, Cho and Dao, Mai Hoang [17] proposed a novel method for detecting APT attacks by leveraging network traffic analysis combined with DL models. They employ individual DL models such as Multilayer Perceptron (MLP), CNN, and LSTM, which were integrated into a combined DL framework to analyze and detect APT attacks in network traffic. The detection process involves two main phases: first, the IP addresses are used to analyze network traffic into flows. The combined DL models are then employed to extract IP features from these flows. Second, using the features extracted in the first phase, the model classifies IPs as either APT attack IP addresses or normal IP addresses. The accuracy ranges between 93% and 98% due to the combination of DL models.

Javed et al. in [18] developed multiple models, including XGBoost, RF, SVM, and AdaBoost for detecting APTs in the IIoT domain. The KDDCup99 [21] dataset was used to assess the performance of these models. The main finding of this study shows that, AdaBoost method outperformed the other techniques, achieving an accuracy of 99.9% with an execution time of 0.012 seconds for APT detection. However, the study did not utilize a specified APT dataset containing distinct APT phases.

In [19], Javed et al. developed a DL model capable of detecting hidden APT attacks in Cyber-Physical Systems (CPSs) integrated with the IIoT in real-time. The authors utilized a Graph Attention Network (GAN) to capture the behavioral features of the attack. Two datasets, DAPT2020 [22] and Edge IIoT, were used to evaluate the study. The results showed that for the DAPT2020 dataset, the model achieved 96.97% accuracy with a prediction time of 20.56 seconds. For the Edge I-IoT dataset, the model achieved 97.5% accuracy with a prediction time of 21.65 seconds.

Dinh-Dong Dau et al. in [20] for improving detection accuracy to APT stages in the SCVIC-APT-2021 dataset , a combination of preprocessing steps is used [22]. These steps included addressing data imbalance using techniques such as resampling, oversampling, undersampling, and cost-sensitive learning. They evaluated the processed data using a range of ML, DL, and ensemble learning algorithms. Among these, XGBoost achieved a Macro F1-score of 95.20%, while LightGBM reached 96.67%, demonstrating significant performance. However, the research did not specify which preprocessing technique improved the detection accuracy.

Table 1, summarizes the reviewed works in terms of the used model, adopted datasets for evaluation, and highlights their limitations.

**Table 1:** Summarized related works

| Work | Proposed system | Dataset | Model | Limitations |
|---|---|---|---|---|
| [12] | Utilizing a DL model with six layers to extract intricate patterns and relationships from network traffic for APT detection. | NSL-KDD | DL, C5.0, and Bayesian | ✓ The study did not utilize a specified APTs dataset.<br>✓ There is no multi-phase detection. |
| [14] | Detecting unknown APTs by analyzing network flow features that traverse the communication channel between the APT attacks and the C&C server. | Contagio blog malware and bigFlows.pcap of TcpPlay | LSTM, CNN, and GBDT | ✓ There is no multi-phase detection.<br>✓ High complexity due to the use of multiple models for feature selection. |
| [15] | Exploiting static, dynamic, and network-related features to distinguish APT-related malware from non-APT malware executables. | Private dataset | RF | ✓ The study did not compare selected APT features with normal traffic. |
| [16] | Classifying APT attacks in IIoT environment by analyzing long attack sequences and extended attack durations using a proposed BERT schema. | Private dataset | LSTM and CNN | ✓ The study did not utilize a specified APTs dataset.<br>✓ High complexity due to the use of two DL models. |
| [17] | Detecting APT by analyzing network traffic flows and extracting IP features using the combined DL models, which are then used to classify IPs as either APT attacks or normal. | CTU-13 | MLP, CNN, and LSTM | ✓ There is no multi-phase detection.<br>✓ High complexity due to the use of multiple DL models.<br>✓ The method does not detect individual APT attacks. |
| [18] | Developing multiple ML models for detecting APTs in the IIoT domain. | KDDCup99 | XGBoost, RF, SVM, and AdaBoost | ✓ The study did not utilize a specified APTs dataset. |
| [19] | Detecting hidden APT attacks in Cyber-Physical Systems (CPSs) integrated with the IIoT in real-time, utilizing a graph attention network to capture the attacks's behavioral features. | DAPT2020 and edge IIoT | GAN | ✓ There is no multi-phase detection. |
| [20] | Improve the detection accuracy for APT stages by applying different preprocessing techniques. | SCVIC-APT-2021 | XGBoost, RF Light GBM | ✓ The research did not specify which preprocessing technique improved the detection accuracy. |

The key issues identified from previous related works are that researchers did not use datasets specifically designated for APT attacks. Additionally, APTs can have unique features that may not be well captured in traditional intrusion detection datasets. Using a dataset tailored to APT scenarios can help ensure the proposed techniques are effective in the specific context of APT detection.

Another limitation is that the reviewed literature did not address multi-phase detection scenarios. As mentioned in the introduction section, APT often involves a series of coordinated phases. The phase at which an attack is detected can significantly affect the mitigation strategy employed by the target organization. For example, if the APT is detected early, such as during the initial compromise, the organization can focus on containment and eradicating the threat by isolating infected systems.

## 3    APT Background

APT is an extended cyber-attack aimed at gaining unauthorized access to a target network or server while remaining undetected for a prolonged period. To achieve this objective, the APT progresses through a series of phases, as illustrated in Fig. 1.

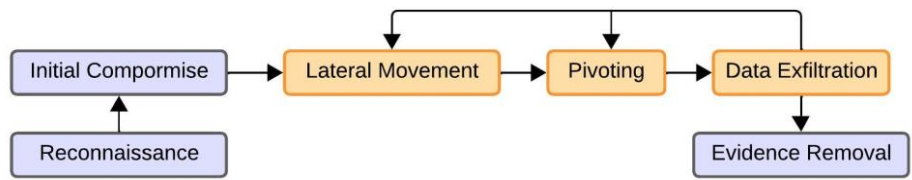

**Fig. 1.** APT phases

- Reconnaissance Phase: In this phase, the APT attacker gathers information about the target to be used in the next phases. Various resources are utilized for this, including details about the organization's size, location, and IT infrastructure. Additionally, information related to the organization's employees, such as decision-makers, access levels, contact details, and social media profiles, may also be collected. Furthermore, information about the organization's network such as IP addresses, network topology, and firewall configurations, may also be gathered [21]. Attackers employ various techniques to accomplish this phase, such as social engineering, where employees of the organization receive malicious information through phishing emails or phone calls. Another technique involves scanning the target system to identify vulnerabilities without triggering any defensive mechanisms.
- Initial Compromise Phase: The attacker selects an appropriate entry point into the target system based on the information identified during reconnaissance. An entry point would include those IoT devices that are the most vulnerable to cyber-attacks since extensive security measures have not yet been implemented due to their limited processing power, normally running on outdated firmware [22]. Early detection of

APT at this stage is crucial for minimizing damage, preventing data breaches in the subsequent phases, and safeguarding the organization's reputation and critical assets. Systems under attack require comprehensive security solutions that address diverse tactics, including IDSs, endpoint protection, network segmentation, and continuous security monitoring.
- Lateral Movement Phase: During this phase, attackers seek to expand their presence and move laterally across the network to compromise additional systems and resources. The primary objective is to explore the network, escalate privileges, and access valuable information. In large companies or organizations, the network environment typically comprises multiple domains or subnetworks.
- Pivoting Phase: During an APT attack, the assailant extends their assault by navigating through different network segments, allowing them to expand their reach beyond the initially compromised system.
- Data Exfiltration Phase: This phase involves transferring data to the attacker's systems, often using the malware's built-in upload functions. In some cases, the attacker used network protocols such as FTP, HTTP and HTTPS [23].
- Evidence Removal Phase: The final phase involves deleting log data and removing malware to cover the attackers' tracks.

In this study, we introduced methods for identifying the key features pertinent to each phase of the APT. These identified features contribute to the development of lightweight IDS systems that can detect APTs at various phases using a minimal set of relevant features. This enables organizations to recognize and respond to threats much earlier in the APT lifecycle. Early detection allows security teams to intervene and contain attacks before they can escalate, potentially averting significant data breaches or system compromises.

## 4 Methodology

This section reviews the algorithms and techniques used in developing the proposed feature selection approach. Section 4.1, elaborates on the ML algorithm where XGBoost is utilized as the classifier for identifying APT phases. Section 4.2 discusses the feature selection methods, focusing on the use of SHAP as a model-agnostic approach used in the development of the proposed method.

### 4.1 XGBoost

Extreme Gradient Boosting (XGBoost) is a popular ML algorithm for regression and classification tasks. It operates on the principles of ensemble learning, by combining predictions from multiple weak learners typically decision trees. The core of XGBoost lies in its objective function, which guides the model parameters during the training process.

The objective function, represented by (1), comprises two components. The first component is the loss function, $l(y_i, \hat{y}_i)$, which measures the difference between the actual value $y_i$ and the predicted value $\hat{y}_i$. The second component is a regularization term, $\Omega(f_k)$, designed to prevent overfitting by penalizing model complexity [24].

$$obj(\theta) = \sum_{i=1}^{n} l(y_i, \hat{y}_i) + \sum_{k=1}^{K} \Omega(f_k) \qquad (1)$$

XGBoost uses features as splitting points during model construction in the training process. The algorithm proposes candidate splitting points for each feature, evaluates these points and selects the best split based on an evaluation score that improves the model's predictions [25].

Reducing the number of features in the dataset results in fewer potential splits to evaluate. This reduction decreases the training time, as the algorithm can iterate more quickly over the dataset and requires less data to store in memory [26][27]. Additionally, minimizing the number of features helps build a more accurate model by focusing on the most important patterns in the data, which can prevent overfitting.

The XGBoost algorithm boasts several key characteristics that make it an ideal choice for developing IDS, where balancing speed and accuracy is crucial. One of its strengths is an advanced tree pruning technique, which stops the construction of a tree when further splitting does not improve the model's performance [28]. This results in a more efficient model with optimal depth, thereby reducing training time and enhancing prediction speed. Furthermore, IDS often contend with imbalanced datasets, where normal traffic significantly surpasses malicious activities. XGBoost effectively addresses this issue by optimizing its performance for skewed data distributions, ensuring accurate identification and classification of attacks even in unbalanced scenarios [29]. These capabilities enable XGBoost to produce accurate and interpretable outcomes with low computational overhead, setting it apart from more complex and resource-intensive approaches like Deep Neural Networks (DNNs).

Also, The combination of XGBoost and the SHAP method (which will be explained in the next section) significantly contributes to faster and more accurate detection. This integration is largely attributed to XGBoost's tree-based framework, which inherently supports efficient SHAP value computation. By utilizing the binary splits within XGBoost's decision trees, SHAP values can be calculated rapidly, providing a clear and efficient assessment of feature contributions. This tree structure not only accelerates the computation process but also enhances the interpretability of the model's predictions, making it an ideal choice for feature selection in IDS.

## 4.2 Features Selection Techniques

Feature selection is the process of identifying the most relevant subset of features that enhance ML performance. This process reduces model complexity, increases efficiency by training with fewer features, and conserves computational resources [30].

There are three feature selection methods: filter-based, embedded methods, and wrapper-based. Filter-based methods select relevant features based on their statistical properties without considering the specific ML algorithm to be used. A drawback of this method is that it does not account for interactions between features. Common filter methods include Correlation Coefficient (C.C), Chi-Square and Analysis of Variance (ANOVA). In embedded methods such as Lasso Regression (LR), relevant features are selected during the training process of the ML model. However, this method is inherently tied to the specific model being used. The third method, wrapper-based, selects features based on the model's performance on various subsets of features, using a greedy search strategy to find the best subset. Although wrapper methods can be computationally expensive, they capture feature interactions and provide accurate predictions. Wrapper-based requires high computational cost because identifying the best feature subset requires training an ML model and measuring its performance for each potential subset [31].

A frequently used technique for implementing the wrapper method is sequential feature selection, which can be performed in a forward or backward manner. In the forward approach, the method iteratively adds the best new feature to the set of selected features in a greedy manner, starting with an empty set and adding features until no further improvement is observed. Conversely, the backward approach begins with all features and iteratively removes the least important ones until no further improvement is achieved [32].

In this work, we utilized the forward selection technique with SHAP values as a feature ranking method. SHAP values explain model predictions by decomposing the output into contributions from each feature. For a given prediction f(x) of an instance x, SHAP uses the Shapley value framework to assign each feature a value representing its contribution to the prediction. The model's prediction can be expressed as the sum of the base value (expected output) and the SHAP values of the features as in (2):

$$f(x) = \emptyset_0 + \sum_{i=1}^{M} \emptyset_i \qquad (2)$$

Where:

- $f(x)$ is the model's prediction for instance $x$,
- $\emptyset_0$ is the base value (the mean model prediction over the entire dataset),
- $M$ is the total number of features,
- $\emptyset_i$ is the SHAP value of feature $i$, representing its contribution to the prediction.

The SHAP value $\emptyset_i$ of feature $i$ is calculated as in (3):

$$\emptyset_i = \sum_{S \subseteq F\backslash\{i\}} \frac{|S|!(|F|-|S|-1)!}{|F|!} [f(S \cup \{i\}) - f(S)] \qquad (3)$$

This formula accounts for all possible subsets $S$ of the feature set $F$, where the term $[f(S \cup \{i\}) - f(S)]$ is the marginal contribution of feature $i$ to the prediction. The weighting factor $\frac{|S|!(|F|-|S|-1)!}{|F|!}$ ensures fair attribution of the feature's contribution by considering the order of features in different subsets.

Thus, the prediction is explained as the base value $\emptyset_0$, adjusted by the SHAP values $\emptyset_i$ of each feature that together explain how features have contributed to the final prediction.

## 5  Proposed Feature Selection Method

Fig. 2 illustrates the workflow of the proposed method, detailing the process for developing a feature selection method aimed at identifying the key features essential for accurate predictions or classifications by the model.

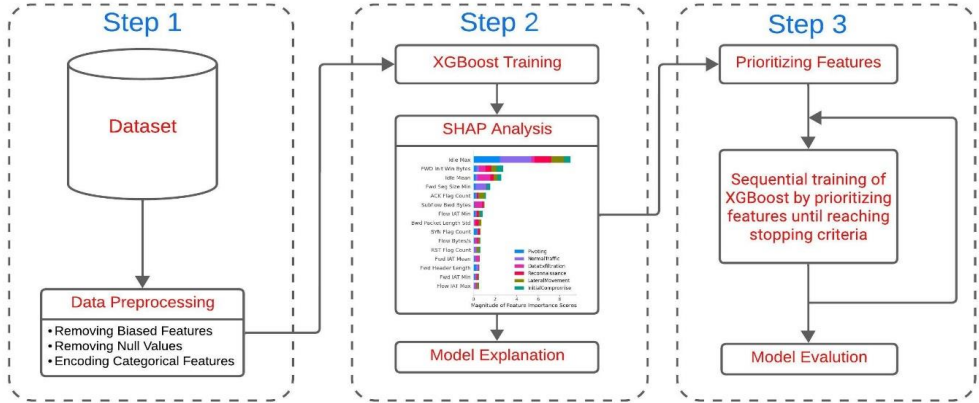

**Fig. 2.** Proposed method pipeline

The pipeline initiates with reading the dataset and preprocessing it to ensure it is appropriately prepared for model training. Next, feature analysis is conducted to obtain feature importance scores, which are utilized to select only the most relevant features. The proposed method is described in the following steps:

- Step 1: The first step of the pipeline involves reading the dataset used to evaluate the proposed method. The IDS dataset comprises features extracted from network traffic flow such as packet size, flow duration, and protocol type using specialized tools like CICFlowMeter and NetFlow. However, not all these features are standardized for ML. So, preprocessing is a very vital step in using them in any ML model. The preprocessing involves the removal of biased features, dealing with null values, and conversion of categorical features into numeric.

- Step 2: In this step, we analyze the features, beginning with training the XGBoost algorithm. Once the model is trained, an explainer object is created to elucidate the inner workings of the trained model. This explainer helps us understand how the model makes predictions based on the input features. We then leverage the SHAP method on the explainer object to analyze the testing set, computing SHAP values for each data point to provide insights into the contribution of each feature to the model's predictions. The output of this step is a list of SHAP features importance scores in decreasing order. Additionally, SHAP analysis in this phase allows us to generate informative plots that help in understanding how individual features influence model predictions.

- Step 3: In this step, we determine the most relevant features using the forward selection technique, functioning as a wrapper feature selection method. However, we do not utilize a greedy search in this selection process. Instead, we commence with an empty feature set, denoted as ($S_{best}$), and add the top feature ($j$) from the SHAP feature importance score list obtained in Step 2. After adding each feature, we evaluate the model's performance, specifically in terms of the macro-averaged F1-score, denoted as ($f_{1_{best}}$). We continue to iteratively add top features until there is no further improvement in the model's performance is observed. This selection strategy allows us to avoid searching through all possible combinations of feature subsets, concentrating only on the top features in the list, thus reducing the computational cost of the selection algorithm. Ultimately, the selected features in the set ($S_{best}$) represent the subset of features that yield the best performance according to the macro averaging F1-score metric. The forward selection process is described in Algorithm 1.

**Algorithm 1:** Feature Selection using SHAP Values and XGBoost

| | |
|---|---|
| 1 | **Input:** SHAP features an importance score list |
| 2 | **Output:** Set of relevant features $S_{best}$ |
| 3 | $f_{1_{best}} = 0$ |
| 4 | $S_{best} = \{\}$ // Initialize an empty set of selected features |
| 5 | **for** $j$ in range of (num-features) **do** |
| 6 |     $S_{new} = S_{best} \cup \{j\}$ // Add top feature $j$ |
| 7 |     Train **XGBoost model** with $S_{new}$ |
| 8 |     $f_{1_{new}}$ = macro averaging F1-score metric (**XGBoost model**) |
| 9 |     **If** $f_{1_{new}} > f_{1_{best}}$ |
| 10 |         $f_{1_{best}} = f_{1_{new}}$ // Update best performance |
| 11 |         $S_{best} = S_{new}$ // Update selected features |
| 12 |     **end** |
| 13 | **end** |

## 6 Experimental Setup

The proposed algorithm was implemented using Python 3.9. The implementation was carried out on a computer equipped with an Intel(R) Core(TM) i5-6300U CPU (2.40GHz) and 8GB of DDR4 RAM. The hyperparameters of XGBoost were carefully tuned to optimize the model's performance during training, as outlined in Table 2.

**Table 2:** Hyperparameter of XGBoost

| Hyperparameter | Description | Value |
|---|---|---|
| n_estimators | Number of trees | 100 |
| learning_rate | Step size shrinkage used to prevent overfitting | 0.3 |
| max_depth | Maximum depth of a tree | 6 |
| min_child_weight | Minimum sum of weights needed in a child node. | 1 |
| objective | Loss function to be minimized | multisoftmax |
| sample_weight | Weights for addressing class imbalance | Varies by dataset |
| gamma | Minimum loss reduction required to make a split | 0 |
| lambda | L2 regularization term to prevent overfitting | 1 |
| alpha | L1 regularization term to prevent overfitting | 0 |

### 6.1 Experimental Description

SCVIC-APT-2021 dataset was used to evaluate the proposed method [33]. This dataset was selected for modeling APT attacks due to several key factors. It contains real APT attacks and covers comprehensive APT phases, as outlined in Table 3, offering a more accurate representation of advanced attack behaviours. Additionally, the dataset consists of 84 features, which are crucial for assessing the effectiveness of our method in reducing the number of features used in the prediction process. Moreover, the dataset not only captures individual APT phases but also provides a holistic view of the interconnectedness of their interconnectedness, forming a robust foundation for research in APT detection.

**Table 3:** Attack techniques applied for each phase

| APT phases | Attack techniques |
|---|---|
| Reconnaissance | Scanning and gathering information for both host and network |
| Initial Compromise | Very Secure FTP Daemon (VSFTPD) |
| Lateral Movement | Pass the Hash/Ticket, Remote Desktop Protocol, WMI |
| Pivoting | AutoRoute, Socks4a, Proxy Chain |
| Data Exfiltration | DNS Tunneling, C2 Tunneling, Encode and Encrypt |

Several preprocessing steps were applied to prepare the data for the ML model. First, specific features such as: 'Flow ID', 'Src IP', 'Src Port', 'Dst IP', 'Dst Port', and 'Timestamp' were removed to prevent the ML model from biasing towards the attacks detection, as these features are used for labeling traffic in the dataset. Next, all samples containing null values were removed, reducing the number of samples from 315,607 to 313,003. Categorical features were then converted into numerical values using the label encoder method. Finally, the dataset was split into two sets: 80% for training and 20%, ensuring the model could be evaluated on unseen data. Table 4 provides the number of samples in the training and testing sets for each class.

As shown in Table 4, the dataset suffers from an imbalanced classification problem, with an unequal distribution of classes in the training set. This imbalance can cause the model to become biased towards the majority class, leading to more frequent but incorrect predictions for that class. Consequently, the model's performance on the minority classes may be significantly worse [32].

**Table 4:** Number of training and testing samples for each class

| Class | No. of samples | | | Total percentage |
|---|---|---|---|---|
| | Training | Testing | Total | |
| Normal Traffic | 246253 | 61564 | 307817 | 98.30% |
| Reconnaissance | 867 | 217 | 1084 | 0.30% |
| Initial Compromise | 120 | 30 | 150 | 0.04% |
| Lateral Movement | 695 | 174 | 869 | 0.20% |
| Pivoting | 1986 | 496 | 2482 | 0.70% |
| Data Exfiltration | 481 | 120 | 601 | 0.19% |

To address this class imbalance, class weights were computed for each class $i$ in the dataset using the following in (4):

$$Class\ Weight_i = \frac{Total\ Samples\ Count}{Number\ of\ classes * Class_i\ Samples\ Count} \quad (4)$$

These computed class weights were then used to set the sample_weight parameter in the XGBoost algorithm, assigning different weights to individual samples. During training, XGBoost fits decision trees to the dataset with these sample weights, giving more importance to samples from minority classes. This adjustment ensures the model focuses on correcting errors made on underrepresented classes. Each decision tree is trained on the weighted dataset, progressively improving the model's performance on challenging examples [34].

### 6.2    Evaluation Metrics

The model's performance is evaluated using several metrics, including accuracy (Acc), precision (Pre), recall (Rec), and F1-score. Accuracy measures the overall correctness of the model's predictions. Precision assesses how many of the positive predictions are actually correct, while recall (also known as the detection rate) quantifies the proportion of actual positive that were correctly identified. The F1-score provides a balanced evaluation by combining both precision and recall into a single metric.

These metrics are derived from four fundamental values: true positives (TP) and true negatives (TN), where the model correctly predicts the positive and negative classes, respectively, and false positives (FP) and false negatives (FN), where the model incorrectly classifies the classes. Standard metrics are calculated as in (5), (6), (7), and (8) [35].

$$\text{Accuracy} = \frac{(TP + TN)}{(FP + FN + TP + TN)} \tag{5}$$

$$\text{Precision} = \frac{TP}{(TP + FP)} \tag{6}$$

$$\text{Recall} = \frac{TP}{(TP + FN)} \tag{7}$$

$$F1 - \text{score} = \frac{2 * (\text{Precision} * \text{Recall})}{(\text{Precision} + \text{Recall})} \tag{8}$$

We also utilize macro-averaging for F1-score, recall, and precision, along with the weighted-averaging F1-score, as both approaches are particularly effective in addressing class imbalance. Macro-averaging calculates each metric F1-score, recall, and precision independently for every class and then averages them, ensuring all classes are treated equally regardless of their frequency in the dataset. In contrast, the weighted F1-score adjusts for class proportions by taking into account the number of true instances in each class. Together, these metrics offer a balanced and comprehensive evaluation of the model's performance across all classes, highlighting its effectiveness and robustness in managing imbalanced datasets.

## 7  Results and Discussion

This section presents the results of the proposed method for detecting different phases of APT. The results of the proposed approach are divided into two subsections. Section 7.1 outlines the results of applying the proposed feature selection method, focusing on its impacts across all classes in the dataset. In Section 7.2, we conduct a comparative analysis to demonstrate the effectiveness of our approach. First, we compare the results of our feature selection method with different feature selection techniques. Next, we assess the overall performance of the developed IDS based on the proposed feature selection method in comparison to various IDS-based ML techniques, such as Random Forest (RF), DNN, and Decision Tree (DT), and compare it with the existing works discussed in Section 2.

### 7.1 Evaluation of the Proposed Feature Selection Method

In this section, we evaluate the effectiveness of the proposed feature selection method, which leverages SHAP to enhance the interpretability and relevance of selected features across different phases APTs. By utilizing SHAP, the method identifies a reduced set of features while providing insights into the contribution and importance of each feature in the detection process. Figure 3 shows SHAP values for different phases, illustrating the significance of various features in identifying them. This transparency helps us understand each feature's impact on the model's predictions. For instance, features like 'Idle Max' and 'Idle Mean' have high SHAP values, indicating their significant influence on decisions. The colorful bars in the figure illustrate the significance of each feature across various APT phases. SHAP values facilitate phase specific analysis by indicating the key features at each phase, thereby boosting the model's detection capabilities. For instance, 'FWD Init Win Bytes' plays a crucial role during the Reconnaissance phase. Similarly, 'Idle Max' is the most critical feature for the Normal Traffic phase, whereas 'Packet Length Std' stands out as the most influential feature in the Data Exfiltration phase.

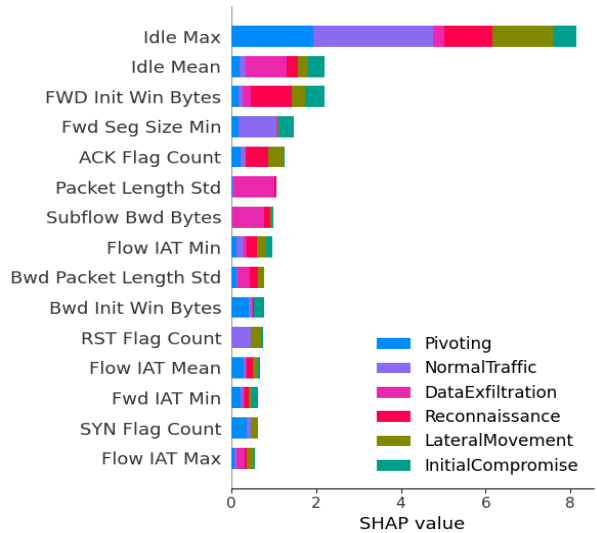

**Fig. 3.** Features importance related to all APT phases

For a more detailed explanation, Fig. 4 shows the top 8 important features for each APT phase. The y-axis lists the feature names, ranked by importance from top to bottom, while the x-axis displays SHAP values. These values assign importance to each feature by calculating its average marginal contribution across all possible feature coalitions. SHAP values can be either positive or negative and are depicted as colored dots. Positive values indicate that the feature contributes to predicting an attack, while negative values suggest contributing to normal traffic. The color represents the feature values, with red indicating high values and blue indicating low ones.

Let us compare two of the most important features in the 'Initial Compromise' phase with the 'Normal Traffic' class. The first feature is 'Idle Max', which refers to a maximum period during which a network connection or communication channel is established but not actively transmitting data. For the 'smileyface' attack in the 'Initial Compromise' phase, red dots are predominantly found to the right of the zero line, indicating that higher idle times increase the likelihood of the model predicting an attack. Conversely, blue dots to the left suggest that lower idle times lead to non-attack predictions. This reflects the attacker's goal of staying undetected as long as possible. In the 'Reconnaissance' phase, attackers may engage in low-profile activities while gathering information about the target network. The 'Idle Max' feature helps detect these subtle activities by highlighting extended periods of inactivity or low-level network interactions that deviate from normal behavior. This is visualized with red dots for all attacks, while the 'Normal Traffic' class is represented by blue dots, indicating lower values consistent with typical communication.

The second feature, 'Fwd Seg Size Min,' represents the smallest segment byte in the forward direction. In the 'Initial Compromise' phase, higher values (red dots) suggest attack traffic, often due to the transmission of sequences of characters (e.g., a smiley face :)) as the username, requiring larger segments. In contrast, 'Normal Traffic', is visualized with blue dots, where smaller values are typical.

For further analysis, Fig. 5 provides an in-depth examination of how the model generates predictions based on individual instances and their feature values. The figure presents examples from two distinct classes: 'Normal Traffic' and 'Reconnaissance.' It highlights the influence of specific features, such as 'Idle Max' on the model's predictions. In the 'Reconnaissance' instance, higher 'Idle Max' values contribute to an attack prediction, while lower values in the 'Normal Traffic' instance lead to a non-attack prediction. This underscores the crucial role of the 'Idle Max' feature in distinguishing between these two classes.

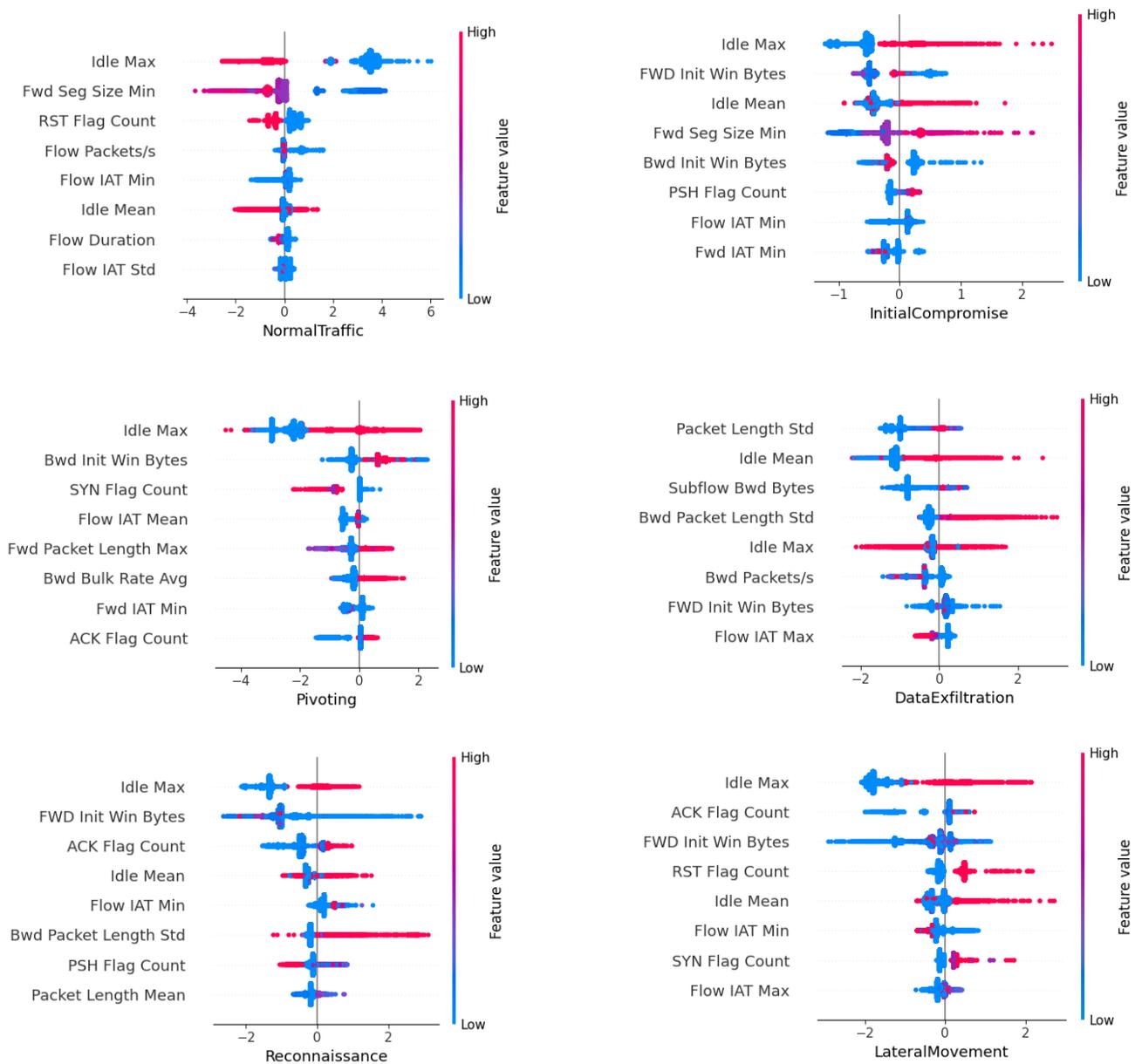

**Fig. 4.** Ordered features importance for each phase in APT

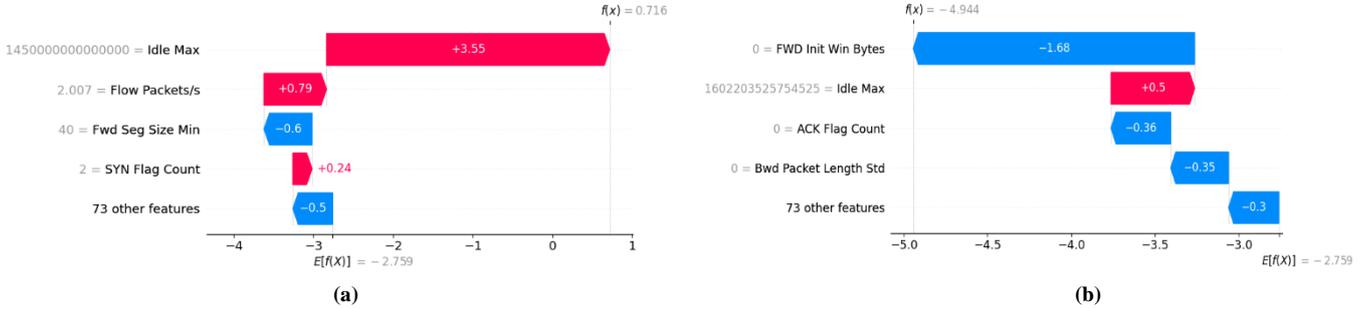

**Fig. 5.** Comparison of feature contributions   (a) 'Normal Traffic'.   (b) 'Reconnaissance'.

Table 5 presents the performance metrics for each phase in the dataset using the proposed method. The model demonstrates high precision across all classes, particularly excelling in accurately identifying 'Normal Traffic' without any errors. This is essential for reducing false alarms and maintaining confidence in the model's predictions.

**Table 5:** Performance metrics for dataset classes

| Classes | Pre | Rec | F1-score |
|---|---|---|---|
| Normal Traffic | 100% | 100% | 100% |
| Reconnaissance | 90% | 92% | 91% |
| Initial Compromise | 100% | 87% | 91% |
| Lateral Movement | 89% | 93% | 91% |
| Pivoting | 96% | 97% | 97% |
| Data Exfiltration | 98% | 87% | 92% |

Notably, the model distinguishes each class with 100% accuracy while utilizing only 12 out of 77 features. This improvement stems from the removal of irrelevant features, leading to a more concise and meaningful data representation that the model can learn from more effectively. Furthermore, the reduction in the number of features significantly decreases the computational cost for both training and prediction.

### 7.2   Comparative Results

To evaluate the performance of SHAP-based feature selection in developing a multiclass classification IDS, we compare it with widely used filter-based feature selection methods, including C.C, Chi2 and ANOVA [28]. In addition to the LR, which is embedded feature selection method. The comparison is made using macro-averaged precision, recall, and F1-score.

As shown in Table 6, the SHAP-based feature selection method outperforms both filter and embedded feature selection methods across all metrics. This superior performance is attributed to SHAP's ability to capture per-feature importance for each sample, other methods only assess the relationship between features and the target label. Additionally, SHAP considers feature relevance through feature interactions, evaluating each feature's contributions in the context of other features, which enables it to capture complex dependencies and interactions between variables.

**Table 6**: Comparison of Different Feature Selection Algorithms

| Method | Relevant 12 Features | Macro-average | | |
|---|---|---|---|---|
| | | Pre | Rec | F1-score |
| C.C | 'Bwd Packet Length Std', 'Bwd Packet Length Max', 'Fwd PSH Flags', 'Bwd IAT Std', 'Packet Length Mean', 'Idle Max', 'Packet Length Std', 'ACK Flag Count', 'Active Mean', 'Active Max', 'Active Min', 'Subflow Fwd Bytes' | 0.94 | 0.89 | 0.91 |
| LR | 'Total Bwd packets', 'Total Length of Fwd Packet', 'Bwd Packet Length Min', 'Bwd Packet Length Std', 'Fwd PSH Flags', 'Packet Length Variance', 'FIN Flag Count', 'SYN Flag Count', 'ACK Flag Count', 'Fwd Seg Size Min', 'Active Mean', 'Active Max', 'Active Min', 'Idle Max' | 0.94 | 0.90 | 0.92 |
| Chi2 | 'Flow Duration', 'Total Fwd Packet', 'Total Length of Fwd Packet','Fwd Packet Length Max', 'Fwd Packet Length Std', 'Fwd IAT Total','Bwd IAT Total', 'Fwd PSH Flags', 'Fwd Header Length','Packet Length Variance', 'ACK Flag Count', 'Fwd Act Data Pkts' | 0.88 | 0.78 | 0.82 |
| ANOVA | 'Total Fwd Packet', 'Total Bwd packets', 'Total Length of Fwd Packet','Fwd Packet Length Max', 'Fwd Header Length', 'Bwd Header Length','Packet Length Max', 'Packet Length Std', 'ACK Flag Count','Bwd Bytes/Bulk Avg', 'Bwd Packet/Bulk Avg', 'Fwd Act Data Pkts' | 0.89 | 0.80 | 0.84 |
| SHAP | 'Bwd Packet Length Std', 'Flow Bytes/s', 'Flow IAT Min', 'Fwd IAT Mean','SYN Flag Count', 'RST Flag Count', 'ACK Flag Count','Subflow Bwd Packets', 'FWD Init Win Bytes', 'Fwd Seg Size Min','Idle Mean', 'Idle Max' | **0.95** | **0.93** | **0.94** |

To further assess the performance of the proposed feature selection, we present a comparison of the performance of different ML-based IDS methods in Table 7. As shown in this table, the proposed XGBoost model with 12 features demonstrates significant efficiency and effectiveness in intrusion detection. While its training time of 67.88 seconds is not the fastest, it is considerably quicker than using 77 features, which takes 189.96 seconds. More importantly, the model excels in detection performance, achieving 100% accuracy, 95% precision, 93% recall, and 94% F1-score. These metrics outperform other methods, including DNN, RF, and DT, which show lower detection performance despite faster training or prediction times. The slight trade-off in training time for the proposed model is justified by its superior detection capabilities, making it an optimal choice for real-world IDSs where accuracy and efficiency are paramount.

**Table 7:** Performance Comparison of Different IDS Methods

| Method | No. of features | Time (seconds) | | Acc | Macro-average | | |
| --- | --- | --- | --- | --- | --- | --- | --- |
| | | Training | Prediction | | Pre | Rec | F1-score |
| Proposed (XGBoost) | 12 | 67.88 | 0.2799 | **100 %** | **95%** | **93%** | **94%** |
| XGBoost | 77 | 189.96 | 0.3280 | **100 %** | 95% | 94% | 94% |
| DNN | 77 | 195.44 | 6.2772 | 99% | 55% | 71% | 59% |
| RF | 77 | 31.13 | 0.7013 | 98% | 57% | 85% | 65% |
| DT | 77 | **2.11** | **0.0180** | 98% | 44% | 76% | 51% |

Table 8 further highlights the proposed model outperforms several other research studies that used XGBoost as a baseline model on the same dataset.

**Table 8:** Compression with XGBoost Algorithm in Other Studies

| Reference | No. of features | Macro-average F1-score | Weighted-average F1-score |
| --- | --- | --- | --- |
| Baseline model [33] | 77 | 70.9% | 99.4% |
| PKI model [36] | 49 | 80.09% | Not addressed |
| Progressive PKI model [36] | 49 | 81.27% | Not addressed |
| XGBoost model [37] | 77 | Not addressed | 97% |
| Proposed model | **12** | **94%** | **100%** |

By selecting only 12 features out of 77, the proposed system significantly reduces computational complexity, ensuring scalability and practicality for deployment in high-speed networks and resource-constrained environments. This feature selection minimizes computational overhead by focusing on the most relevant data, enabling efficient detection with limited resources. The streamlined feature set reduces data storage requirements, simplifies analytical processes, and enhances system responsiveness. Consequently, the IDS achieves faster threat detection, allowing security teams to respond promptly and mitigate potential attacks effectively.

## 8 Conclusion

This study addresses the escalating threat of Advanced Persistent Threats (APTs) by developing a robust and resource-efficient Intrusion Detection System (IDS). This system is built on a proposed feature selection method, where the XGBoost ML algorithm is integrated with Explainable AI (XAI) to select a minimal set of influential features for detecting APTs across different phases. The proposed method identifies the most relevant features for each phase separately. Results from the SCVIC-APT-2021 dataset highlight the effectiveness of the proposed method, showing significant improvements in both the macro-average F1-score and recall, reaching 94% and 93%, respectively, while reducing the number of relevant features from 77 to only 12. Furthermore, the method significantly reduces the training and prediction time without compromising detection accuracy.

This research further advances the field by introducing a new ML model that will improve the detection of APTs and their interpretability due to feature selection, making the network hardened against advanced cyber-attacks. As future work, we would conduct experiments on the method for several APT-related datasets to further evaluate its effectiveness.